\documentclass[12pt]{article}
\usepackage{lnfprep}
\usepackage{epsfig}
\usepackage{amsmath}
\newcommand{\be}{\begin{equation}}
\newcommand{\ee}{\end{equation}}
\def\simge{\mathrel{%
      \rlap{\raise 0.511ex \hbox{$>$}}{\lower 0.511ex \hbox{$\sim$}}}}
\def\simle{\mathrel{
      \rlap{\raise 0.511ex \hbox{$<$}}{\lower 0.511ex \hbox{$\sim$}}}}

\begin{document}
\begin{titlepage}
\title{ 
  {\large \bf The KLASH Proposal}\\{\it Axion Calling}
}
\author{
   D. Alesini$^1$, D. Babusci$^1$, D. Di Gioacchino$^1$, C. Gatti$^1$, 
   G. Lamanna$^2$, C. Ligi$^1$\\
{\it ${}^{1)}$  INFN, Laboratori Nazionali di Frascati}\\
{\it ${}^{2)}$ Universit\`a di Pisa and INFN} } 
\maketitle
\baselineskip=14pt

\begin{abstract}
We propose a search of galactic axions with mass about 0.2~$\mu$eV using a large volume resonant cavity, about 50~m$^3$, cooled down to 4~K and immersed in a moderate axial magnetic field of about 0.6~T generated inside the superconducting magnet of the KLOE experiment~\cite{KLOETDR} located at the National Laboratory of Frascati of INFN. This experiment, called KLASH (KLoe magnet for Axion SearcH) in the following, has a potential sensitivity on the axion-to-photon coupling, $g_{a\gamma\gamma}$, of about $6\times10^{-17}$ $\mbox{GeV}^{-1}$, reaching the region predicted by KSVZ~\cite{KSVZ} and DFSZ~\cite{DFSZ} models of QCD axions.     
\end{abstract}

\vspace*{\stretch{2}}
\begin{flushleft}
\end{flushleft}

\end{titlepage}
\pagestyle{plain}
\setcounter{page}2
\baselineskip=17pt
\section{Introduction}
\label{sec:introduction}
The axion is a pseudoscalar particle predicted by S.~Weinberg~\cite{Weinberg} and F.~Wilczek~\cite{Wilczek} as a consequence of the mechanism introduced by R.D.~Peccei and H.~Quinn~\cite{PecceiQuinn} to solve the ``strong CP problem''. 
Axions with a mass in the $\mu$eV range are also well motivated dark matter candidates. The search of galactic-axion with a "Haloscope'', a resonant cavity immersed in a strong magnetic field 
as proposed by P.~Sikivie~\cite{Sikivie}, is a well established technique. Nowadays, a single experiment reached the sensitivity to probe the existence of QCD axions in the sub-meV region, 
ADMX~\cite{ADMX}. By using as haloscope composed of a resonant cavity with volume 0.2~m$^3$ and unloaded quality factor about $2\times 10^5$ immersed in a magnetic field of 7.6~T,  
ADMX probed the QCD-axions in the mass range between 2 and 3 $\mu$eV. 

Assuming the local dark matter density, $\rho_a=0.45$~GeV/cm$^3$, is due to axions and the cavity frequency $\nu_c$ exactly equal to the axion mass $m_a$, the conversion power inside an 
haloscope is~\cite{YWL}:
\begin{equation}
\label{eq:power}
P_{\mbox{sig}}=\left( g_{\gamma}^2\frac{\alpha^2}{\pi^2}\frac{\hbar^3 c^3\rho_a}{\Lambda^4} \right) \times 
\left( \frac{\beta}{1+\beta} \omega_c \frac{1}{\mu_0} B_0^2 V C_{mnl} Q_L \right)
\end{equation}
where, $\alpha$ is the fine-structure constant, $\mu_0$ the vacuum permeability, $\Lambda=87$~MeV is a scale parameter related to hadronic physics, $g_{\gamma}$ the photon-axion coupling 
constant equal to $-0.97(0.36)$ in the KSVZ (DFSZ) model. It is related to the coupling appearing in the Lagrangian $g_{a\gamma\gamma}=(g_{\gamma}\alpha/\pi\Lambda^2)m_a$. The second 
parentheses contain the magnetic field strenght $B_0$, the cavity volume $V$, its angular frequency $\omega_c=2\pi\nu_c$, the coupling between cavity and receiver $\beta$ and the loaded 
quality factor $Q_L=Q_0/(1+\beta)$, where $Q_0$ is the unloaded quality factor. $C_{mnl}\simeq 0.5$ is a geometrical factor depending on the cavity mode.

Since $P_{\mbox{sig}}$ can be as low as $10^{-22}$~W, the cavity is cooled to cryogenic temperatures and ultra low noise cryogenic amplifiers are needed for the first stage amplification. According 
to the Dicke radiometer equation~\cite{Dicke}, the signal to noise ratio $SNR$ is given by:
\begin{equation}
  \label{eq:snr}
  SNR=\frac{P_{\mbox{sig}}}{k_B T_{sys}}\sqrt{\frac{\tau}{\Delta\nu_a}}
\end{equation}
where $k_B$ is the Boltzman constant, $T_{sys}$ is the combination of amplifier and thermal noise, $\tau$ is the integration time and $\Delta\nu_a$ the intrinsic bandwith of the galactic axion signal 
($\Delta\nu_a/\nu_a\simeq10^{-6}$). 
    
\section{The KLOE detector}
\label{sec:kloe}
\begin{figure}[htbp]
  \begin{center}
    \includegraphics[totalheight=7.5cm]{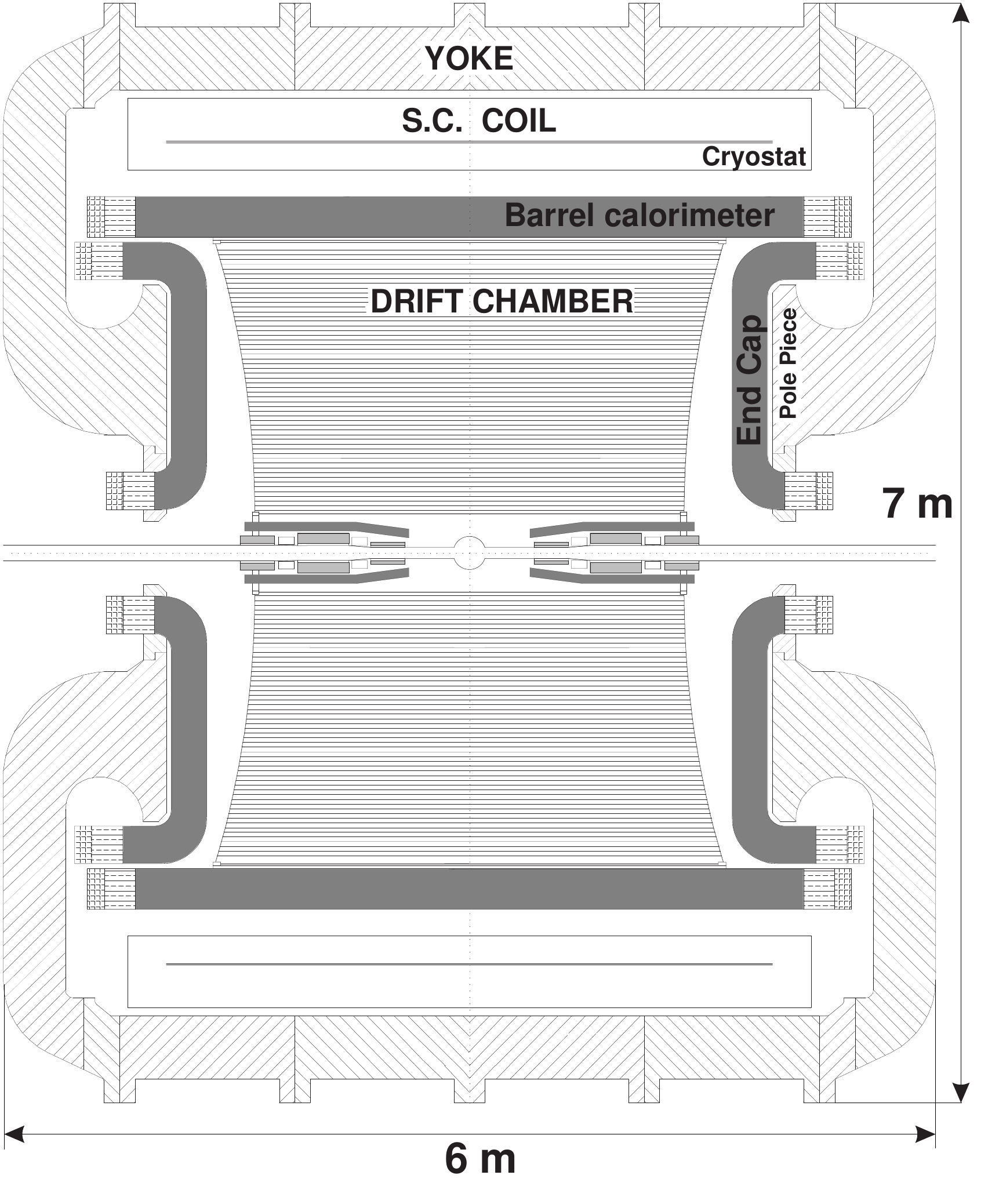}
    \caption{Section of the KLOE detector.}
    \label{fig:kloesec}
  \end{center}
\end{figure}

The KLOE experiment has been recording $e^{+} e^{-}$ collisions at DA$\Phi$NE, the $\phi$-factory at the Laboratori Nazionali di Frascati (LNF), since April 1999. The detector (see Figure~\ref{fig:kloesec}) 
was designed for the study of kaon decays, mainly $K \to \pi^{+} \pi^{-}$ and $K \to \pi^{0} \pi^{0}$~\cite{KLOEreview}.
It  is composed of a cylindrical drift chamber (DC), surrounded by an electromagnetic calorimeter (EmC), both of which are embedded in a magnetic field of 0.52 Tesla. The KLOE detector can be 
moved outside the DA$\Phi$NE collision region and parked inside the adiacent KLOE Experimental Hall.

\section{The KLOE Superconducting Magnet}\label{sec:magnet}
\begin{figure}[htbp]
  \begin{center}
    \includegraphics[totalheight=9cm]{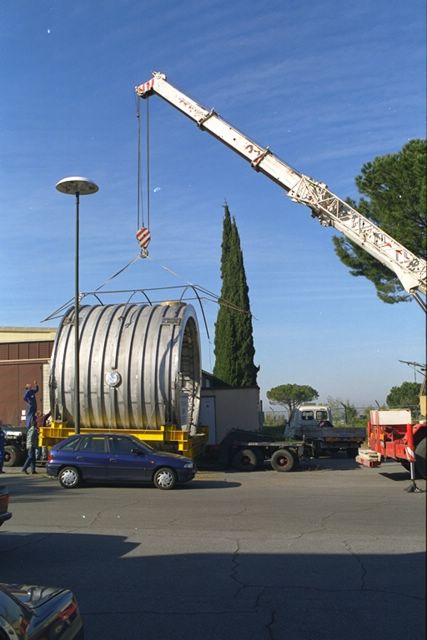}
    \caption{Arrival of the KLOE magnet at LNF in April 1997.}
    \label{fig:kloemag}
  \end{center}
\end{figure}
The KLOE Magnet~\cite{KLOEMAG,MODENA} is an iron shielded solenoid coil made from an aluminium-stabilised niobium titanium superconductor. The coil is cooled and kept at 4.5 K by liquid 
helium (LHe) flowing with a thermo-siphoning cooling method. The cryostat possesses a service turret with a valve box, a Joule-Thomson (JT) valve and a LHe reservoir of $\sim$150 liters. 
5.2 K/3 bar supercritical helium coming from a Linde TCF50-based cryogenic plant enters in the KLOE service turret and liquefies through the JT valve into its reservoir. The LHe then cools the coil 
through the thermosyphon. The coil is surrounded by 2 radiation shields (one on the outer and the other on the inner side) which are cooled with 70~K gaseous helium (GHe) from the cryogenic plant. 
The coil current leads are cooled with LHe from the reservoir. The magnet is under ``continuous cooling'', and the helium flow coming from the cryogenic plant is regulated by a PC with a dedicated 
{\tt Labview} software. 
\vspace*{0.5cm}

In order to run the KLOE magnet in the KLOE Experimental Hall, some modification to the cryogenic plant layout are needed. The main are:
\begin{itemize}
\item move the Linde Valve Box to the KLOE Hall,
\item modify the transfer lines between Cold Box and Valve Box to reach the new Valve Box position,
\item modify the transfer lines between Valve Box and KLOE (it would be sufficient to cut the existing lines),
\item build new transfer lines between the Valve Box and the cavity service turret (see later).
\end{itemize}

Main characteristics and operating parameters of KLOE magnet are summarized in Table \ref{tab:magnet}.
\begin{table}[!ht]
  \begin{center}
    \caption{The KLOE magnet.}
  \label{tab:magnet}
  \vspace*{0.5cm}
    \begin{tabular}{c|c}
      \hline
      \multicolumn{2}{c}{KLOE magnet coil main parameters}    \\ \hline
       Nominal Magnetic Field  & 0.6~T  \\
       Stored Energy   & 14.32~MJ  \\
       Nominal Current   & 2902~A  \\
       Cold Mass   & $\sim$8500~kg  \\
     \hline\hline
      \multicolumn{2}{c}{Nominal steady state heat loads}    \\ \hline
        Magnet Coil & 55 W @ 4.5 K \\
        Current Leads & 0.6 g/s (LHe) \\
        Thermal Radiation Shields & 530 W @ 70 K \\
     \hline\hline
      \multicolumn{2}{c}{Cryostat dimensions}    \\ \hline
       Outer Diameter  & 5.76~m  \\
       Inner Diameter  & 4.86~m  \\
       Overall Length   & 4.40~m  \\
     \hline\hline
    \end{tabular}
  \end{center}
\end{table}

\section{The Resonant Cavity}
\label{sec:cavity}
\begin{figure}[!ht]
  \begin{center}
    \includegraphics[totalheight=7.5cm]{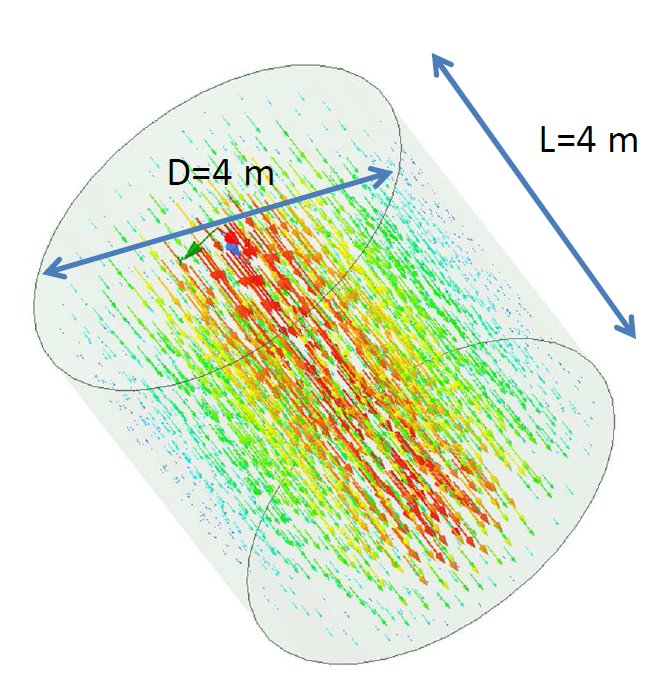}
    \caption{Sketch of the resonant cavity with electric field of the working mode TM$_{010}$.}
    \label{fig:cavity1}
  \end{center}
\end{figure}

\begin{figure}[htbp]
  \begin{center}
    \includegraphics[totalheight=7.5cm]{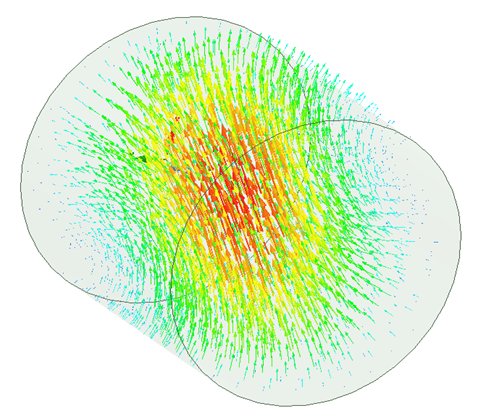}
    \caption{Electric field of the nearest mode TE$_{111}$.}
    \label{fig:cavity2}
  \end{center}
\end{figure}

\begin{figure}[htbp]
  \begin{center}
    \includegraphics[totalheight=7.5cm]{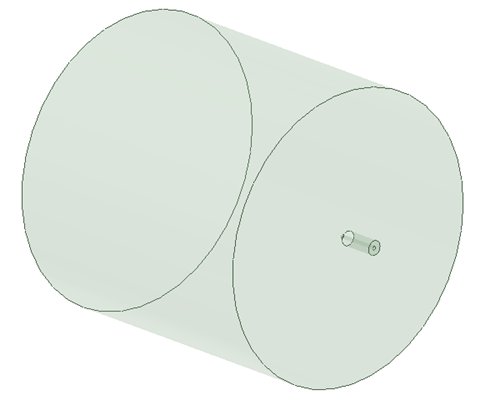}
    \caption{Coaxial cable to probe the signal on the cavity.}
    \label{fig:cavity3}
  \end{center}
\end{figure}

The sketch of the resonant cavity is shown in Figure~\ref{fig:cavity1} with its dimensions. The resonant working mode is the TM$_{010}$ whose electric field configuration is also shown in Figure~\ref{fig:cavity1}. 
The nearest mode is the TE$_{111}$ shown in Figure~\ref{fig:cavity2}. 
The internal surface of the cavity is in copper in order to increase the quality factor of the mode. The resonant frequency and the quality factor, for $T$ = 300 K, of the working mode and of the nearest mode are 
given in Table~\ref{tab:cavity} for two different values of the cavity length $L$. The resonant mode can be coupled to the detector through a coaxial probe as sketched in Figure~\ref{fig:cavity3}. 
\begin{table}[!h]
  \begin{center}
    \caption{Resonant frequency and quality factor of the working mode (TM$_{010}$) and of the nearest mode (TE$_{111}$).}
    \label{tab:cavity}
  \vspace*{0.5cm}
    \begin{tabular}{c|c|c|c|c|}
            & \multicolumn{2}{c|}{$L$ = 4 m}        & \multicolumn{2}{c|}{$L$ = 3.5 m}        \\ [5pt] 
            \hline
       Mode & Frequency [MHz] & $Q_0$ [$T$ = 300 K] & Frequency [MHz] & $Q_0$ [$T$ = 300 K] \\ 
       TM$_{010}$&  57.37    & 153,000  & 57.37  & 146,000  \\ \hline 
       TE$_{111}$&  57.74    & 162,000  & 61.35  & 159,000 \\ 
     \hline\hline
    \end{tabular}
  \end{center}
\end{table}

\begin{figure}[htbp]
  \begin{center}
    \includegraphics[totalheight=7.5cm]{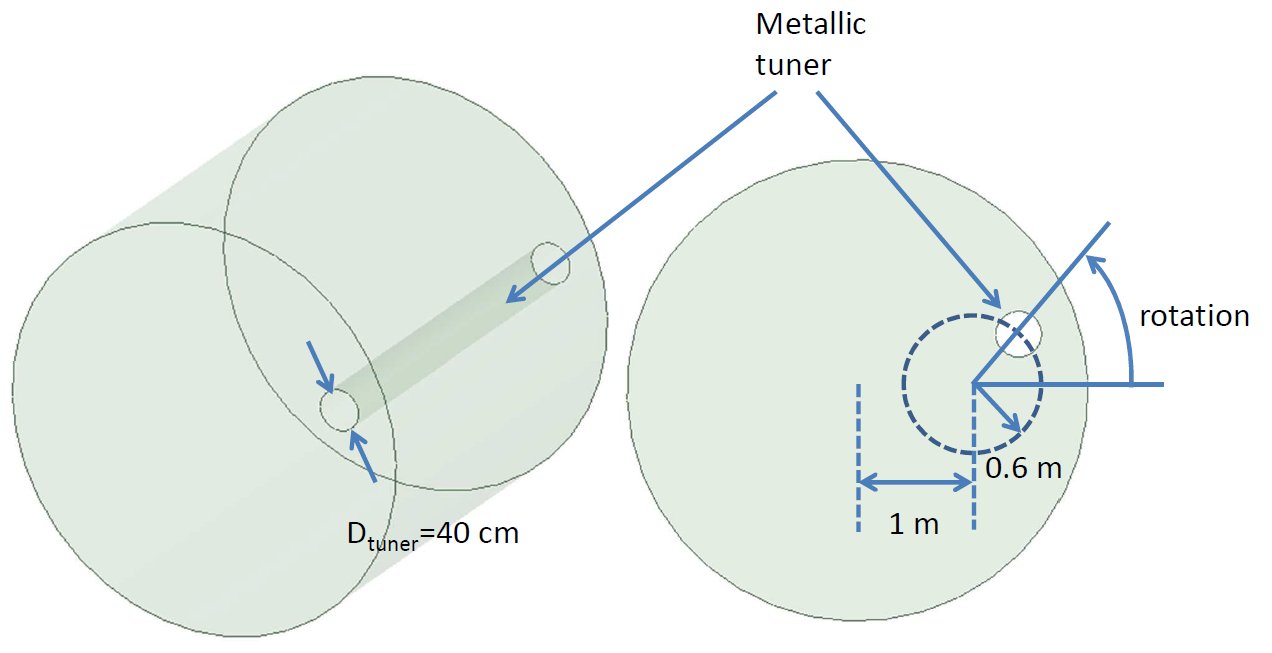}
    \caption{Metallic tuner to change the resonant frequency of the cavity.}
    \label{fig:cavity4}
  \end{center}
\end{figure}

\begin{figure}[htbp]
  \begin{center}
    \includegraphics[totalheight=6.5cm]{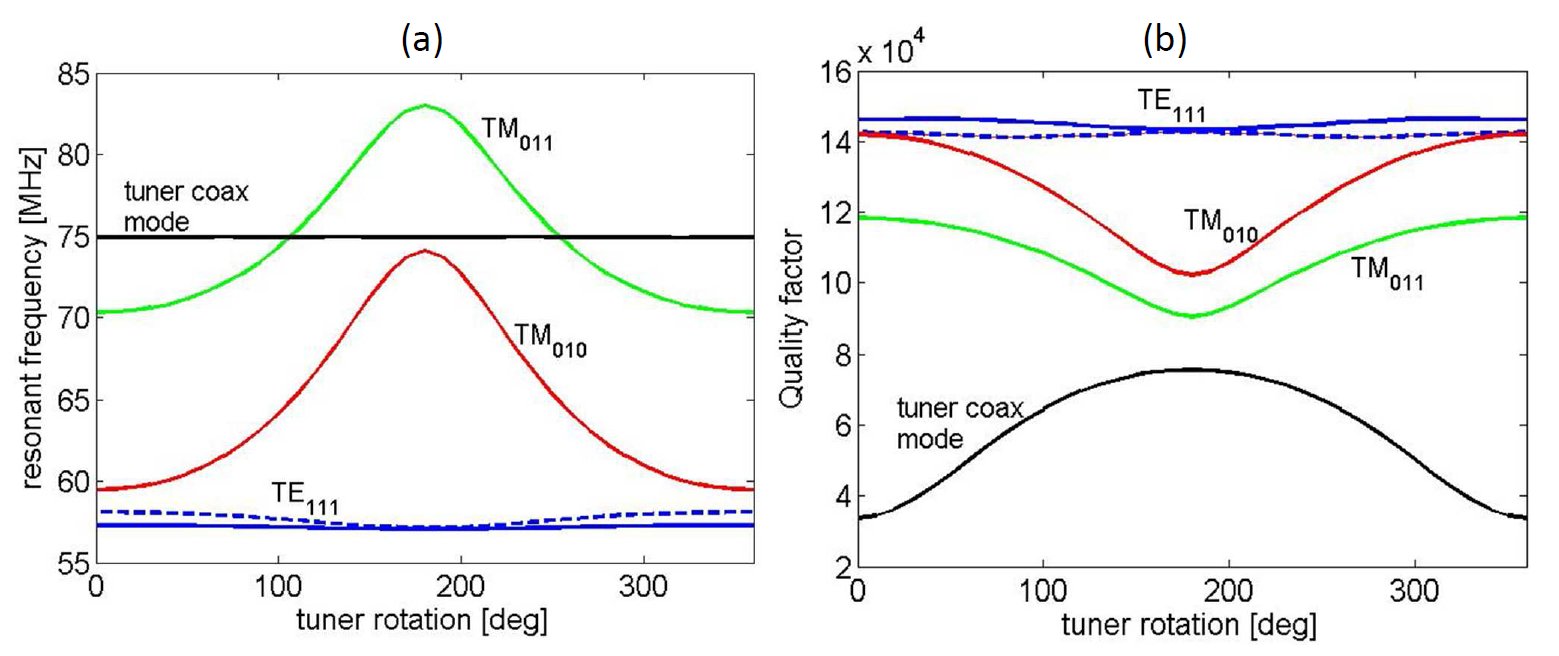}
    \caption{Resonant frequencies and quality factors, at 300 K, of cavity resonant modes as a function of the tuner rotation. By cooling the cavity to 4~K the quality factor is expected to increase 
    considerably.}
    \label{fig:cavity5}
  \end{center}
\end{figure}

Different techniques can be used to change the resonant frequency. One possibility to be investigated is that used in the first paper of Ref. \cite{YWL}. In this case one or more metallic cylinders can be inserted off-axis 
in the cavity and can be rotated sampling different magnetic and electric field regions thus changing the resonant frequency as given in Figure~\ref{fig:cavity4}. As an example the resonant frequencies and the quality 
factors of the cavity resonant modes are given in Figure~\ref{fig:cavity5} for the tuner geometry of Figure~\ref{fig:cavity4}. The insertion of the tuning system changes, obviously, the resonant frequencies and the quality 
factors of the working mode and of the other modes. The cavity volume should be hydraulically connected to its cryostat to allow the air pumping before the cooling.

\begin{figure}[htbp]
  \begin{center}
    \includegraphics[totalheight=14cm]{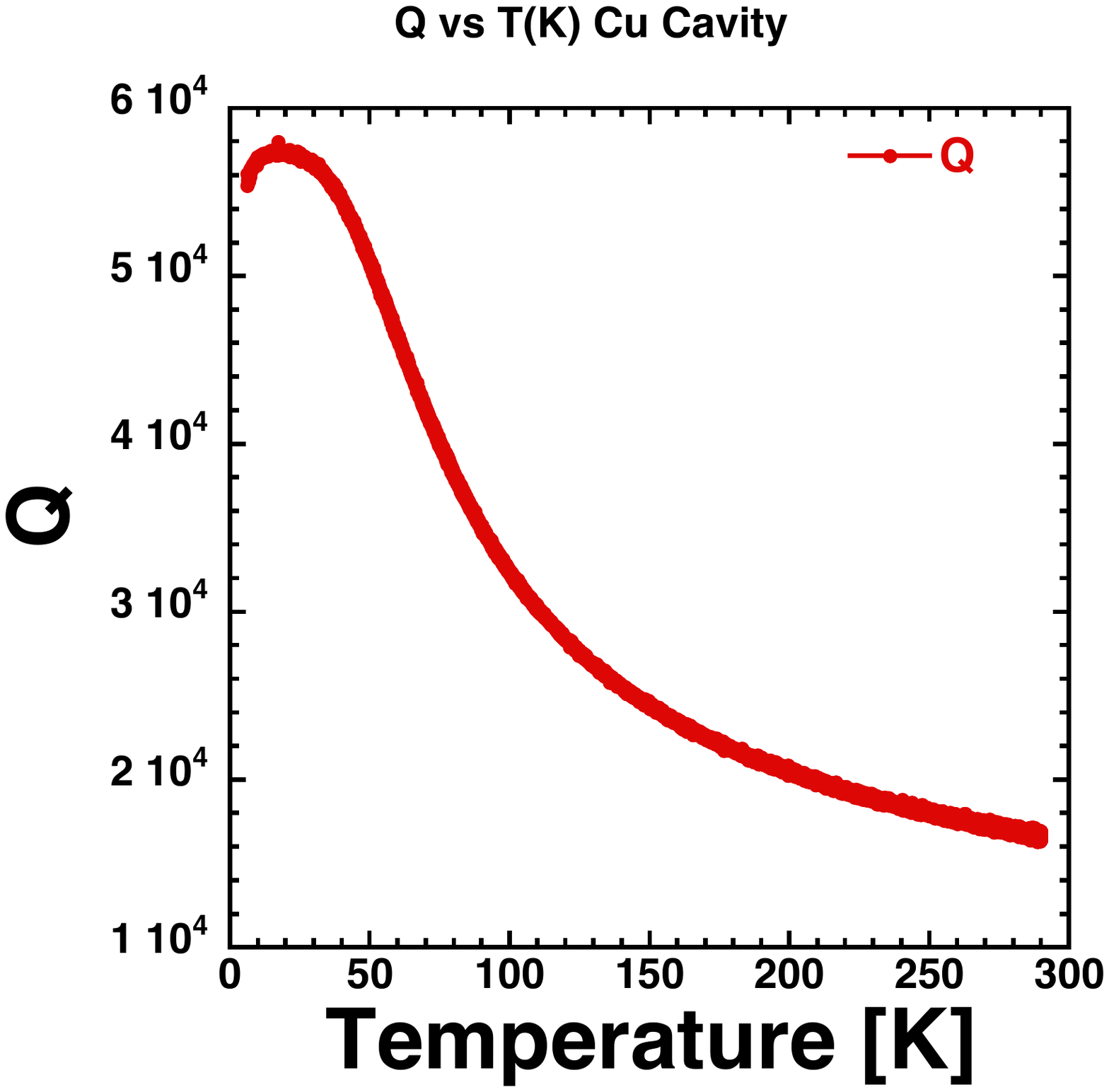}
    \caption{Quality factor as a function of temperature as measured in a copper cylindrical cavity with mode of frequency 14 GHz, during the QUAX R\&D\cite{QUAX}}
    \label{fig:QvsT}
  \end{center}
\end{figure}

By cooling the cavity to 4~K the quality factor is expected to increase considerably. The $Q$ factor of a copper resonant cavity was recently measured at LNF as a function of temperature for a mode of frequency 
14 GHz (Figure~\ref{fig:QvsT}). The $Q$ factor increases by a factor 3 before reaching a plateau at a temperature of about 30~K. Here the anomalous skin depth effect limits further improvements. Since in the 
anomalous limit the surface resistivity scales as $\omega^{2/3}$, for $\omega=57$~MHz we expect to observe stronger improvements. However, in the following we assume a conservative factor 3 improvement, 
corresponding to $Q(4~\mbox{K})$ = 450,000. Moreover, considerations about the natural width of galactic axions ($\delta\omega_{a}/\omega_{a}\sim 10^{-6}$) and on the scan rate, suggest to keep the quality 
factor within $10^6$. The parameters used for the calculation of the KLASH sensitivity are shown in Table~\ref{tab:cavity2}.
\begin{table}[h!]
  \begin{center}
    \caption{Resonant cavity central parameters used in the calculation of the experiment sensitivity.}
    \label{tab:cavity2}
  \vspace*{0.5cm}
    \begin{tabular}{c|c}
      Parameter & Value \\\hline
     $L$ [m] & 4 \\
     $R$ [m] & 2 \\
     Inner surface material & Oxygen Free Copper \\
     Mode & TM$_{010}$ \\
     frequency $\nu_c$ & 57 MHz \\
     $Q_0$ [$T$ = 4 K] & 450,000 \\
     \hline\hline
    \end{tabular}
  \end{center}
\end{table}

\section{Cavity Cryostat}\label{sec:cryostat}
In order to run at 4 K, the resonant cavity will be hosted in a dedicated cryostat, designed to be operated with the same principle of operation of KLOE. A vacuum chamber will surround the cavity, and one 
radiation shield taken at $\sim$ 70 K will be put in the middle. The cold helium will be supplied by the same cryogenic plant operating for KLOE. This can be achieved because the plant has an extra cooling 
capacity which has been formerly dedicated to the FINUDA experiment and the four compensating solenoids operating in DAFNE~\cite{MODENA}. The nominal cooling capacity available for the cavity is a 
liquefaction rate of 0.54 g/s and a refrigeration capacity of 44 W at 4.5 K and 270 W at 70 K, respectively. A service turret, adjacent to the cryostat, will receive both the 5 K and 70 K helium transfer lines from 
the cryogenic plant and will supply the 4.5 K LHe to the cavity and the 70 K to the shield. The LHe will flow in a tube in thermal contact with the cavity. A similar solution is foreseen for the GHe for the shield.

The cavity can be supplied with helium in continuous cooling, as in KLOE. The flow regulation will be provided by a motorized valve controlled by a PC with a {\tt Labview} software. The same program may 
be used for the visualization of all the cavity cryogenic parameters (temperature, pressures, helium level meter, flow meters etc.). In order to allow the vacuum pumping, extreme care must be taken in the 
cryostat design to avoid the walls shrinking due to the differential pressure.

\section{Ultra Low Noise Cryogenic Amplifier}\label{sec:amplifier}
The choice of suitable cryogenic amplifiers is very important because the noise introduced by the front-end electronics chain is, ultimately, the main limitation to the experiment sensitivity.

In the last years several techniques have been developed to improve the amplifiers noise temperature in a wide range of frequencies. These developments, triggered by radio astronomy research and high 
bandwidth space communication, can be exploited for our application.  Among different devices available, the most interesting in our range of frequencies are SiGe HBT (Silicon Germanium Hetero-junction 
Bipolar Transistors) \cite{Bardin,Ivanov}, HEMT (High Electron Mobility Transistor also known as HFET Hetero-junction Field Effect Transistor) \cite{Pospieszalski} and SQUID (Superconducting QUantum 
Interference Device) \cite{Jaklevic}. For all of them has been proved that a noise temperature well below 2~K can be achieved.  While in FET-Based amplifiers design the noise can not decrease below a given 
value, due to the power dissipation in the FET channel, in the SQUID amplifier the quantum limit for the noise can be achieved at very low temperature.

For this reason a scheme with a moderated gain SQUID pre-amplifier and a HFET based amplifier, is suitable for our application, as already proven at higher frequencies \cite{Asztalos:2011bm}. The advantage 
of this scheme is to exploit a very low noise pre-amplifier with moderate gain ($\sim 10-20 $ dB) followed by an amplification stage ($\sim 30$ dB) to obtain a useful signal for readout outside the cryogenic cavity. 
Below 100 MHz, the capability of dc-SQUID as amplifier as been already proven \cite{Hilbert} and the main results are summarized in Table \ref{ref:res}. With a tuned amplifier at 4.2 K a noise temperature of 1.4 K 
was reached. Housing the dc-SQUID in a cryostat at 200 mK noise temperature can be reduced to few hundred mK. 
\begin{table}
\begin{center}
\caption{Results achieved in \cite{Hilbert} for dc-SQUID at cryogenic temperature.}
  \vspace*{0.5cm}
\begin{tabular}{c|c|c|c|}
Temperature & Frequency (MHz)& Gain (dB) & Noise Temperature (K) \\[5pt]
\hline\hline
$T$ = 1.5 K & $\nu = 60$  & $24.0\pm0.5$ & $1.2\pm0.3$ \\
 untuned      & $\nu = 80$ & $21.5\pm0.5$ & $0.9\pm0.3$ \\
                    & $\nu = 100$  & $19.5\pm0.5$ & $1.0\pm0.4$ \\
\hline  
$T$ = 4.2 K & $\nu = 60$ & $20.0\pm0.5$ & $4.5\pm0.6$ \\
   untuned    & $\nu = 80$  & $18.0\pm0.5$ & $4.1\pm0.7$ \\
                    & $\nu=100$  & $16.5\pm0.5$ & $3.8\pm0.9$ \\
\hline
$ T$ = 4.2 K & $\nu = 93$  & $18.6\pm0.5$ & $1.7\pm0.5$ \\
 tuned &                &              &             \\
\hline\hline
\end{tabular}
\label{ref:res}
\end{center}
\end{table}

In addition, since the SQUID can not work properly even with moderate magnetic fields, particular care should be devoted to control intensity and stability of the magnetic field in the region where the SQUID is 
housed. Good SQUID performance and stability of characteristics during the measurements will require the device to be mounted inside a magnetic shield of Nb-Nb3Sn. 

A low noise amplifier stage could be also foreseen at room temperature, outside the cryogenic cavity, before the mixer used for the readout.

\section{The KLASH Sensitivity}\label{sec:sensitivity}
According to Eq. (\ref{eq:power}), the power generated by galactic axions is proportional to the stored magnetic energy times the product of the loaded $Q$ factor and the cavity angular frequency. Comparing 
the factor $F = \omega_{c}B^2VQ_{L}$ for KLASH, ADMX~\cite{ADMX} and YWL~\cite{YWL} we have, approximately, $10^{15}$ rad$\cdot$T$\cdot$m$^3$/s, $4\times10^{15}$ rad$\cdot$T$\cdot$m$^3$/s 
and $5\times10^{13}$ rad$\cdot$T$\cdot$m$^3$/s, respectively.  Therefore, even with a moderate field strength, with a  total volume, about  50~m$^3$, more than two orders of magnitude larger than the ADMX 
volume, 0.2~m$^3$, the KLASH experiment has potentially the sensitivity required to observe galactic axions. Inserting the parameters of Tables~\ref{tab:cavity2} and~\ref{tab:sensitivity}, we reach a sensitivity, 
at 90\% c.l., on the coupling $g_{a\gamma\gamma}$ of  $6\times10^{-17}$ GeV$^{-1}$. \begin{table}[h!]
  \begin{center}
    \caption{The KLASH discovery potential for KSVZ axions. $\beta$ is chosen equal to 2 to optimize the scan rate~\cite{YWL}.}
    \label{tab:sensitivity}
  \vspace*{0.5cm}
    \begin{tabular}{c|c}
      Parameter & Value \\\hline
      $m_a$ [$\mu$eV] & 0.23\\
      $g_{a\gamma\gamma}^{KSVZ}$ [GeV$^{-1}$] & $8.7\times10^{-17}$  \\
      $P_{\mbox{sig}}$ [W] & $1.6\times10^{-22}$  \\
      Rate [Hz] & 3,100 \\
      L$_{cavity}$ [m] & 4 \\
      R$_{cavity}$ [m] & 2 \\
      $B_{max}$ [T]  & 0.6 \\      
      $\beta$ & 2 \\
      $\tau$ [min] & 5\\
      $T_{sys}$ [K] & 4 \\ 
      SNR & 5 \\
      $g_{a\gamma\gamma}$ 90\% c.l. [GeV$^{-1}$] & $6\times10^{-17}$  \\
      \hline\hline
    \end{tabular}
  \end{center}
\end{table}

\begin{figure}[htbp]
  \begin{center}
    \includegraphics[totalheight=11cm]{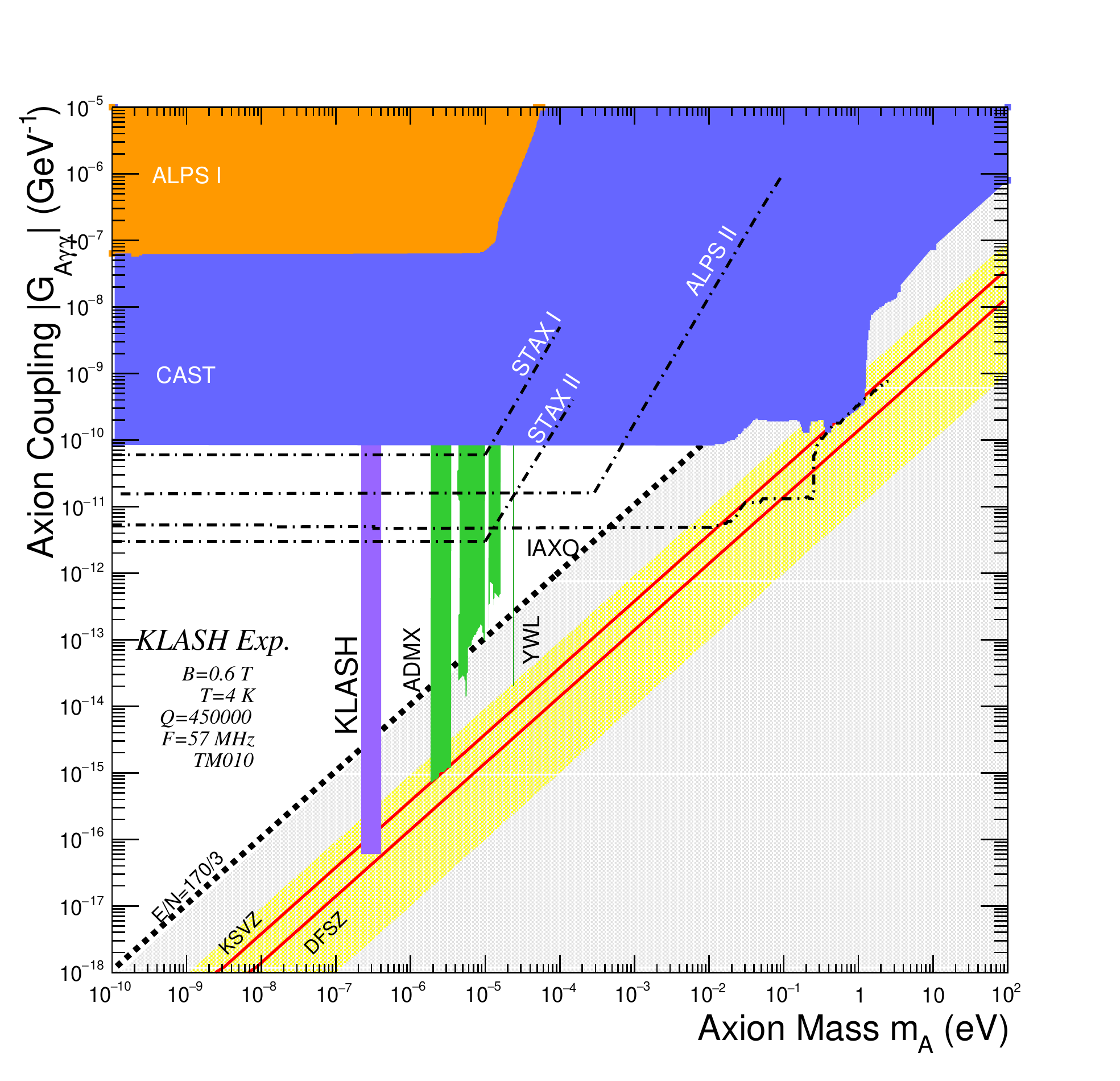}
    \caption{The KLASH discovery potential for axion mass $m_a$ = 0.23 $\mu$eV, 20 MHz bandwith (0.08 $\mu$eV) in one year of data taking and integration time $\tau = 5$~ min. 
    The purple band corresponds to two years of data taking. The yellow band is the preferred region for the KSVZ and DFSZ models. The gray region below the dashed line refers to the axion models discussed in~\cite{NARDI}. Also shown the expected sensitivity of STAX~\cite{STAX}, IAXO~\cite{IAXO} and ALPSII~\cite{ALPS2}.}
    \label{fig:sensitivity}
  \end{center}
\end{figure}

The discovery potential in the coupling-mass plane is shown in Figure~\ref{fig:sensitivity}. The sensitivity band reaches the band predicted for QCD axions of the KSVZ abd DFSZ models.
We considered here an integration time of 5 minutes for a single measurement, after which the cavity is  tuned to a different frequency by means of the tuning rods. The bandwith span with a single measurement 
is equal to $\nu_c/Q_L = 380$~Hz. This corresponds to a total scan bandwith of about 40 MHz in one year of data taking, assuming maximal efficiency, and corresponding to a mass scan of about 0.16~$\mu$eV. 
In section~\ref{sec:cavity} we showed that a single tuning rod increases the resonant frequency of about 15 MHz. With two tuning rods we expect to scan in two years of data taking a region of about 30-40 MHz.

\newpage 
\section{Acknowledgements}
We would like to thank the KLOE Collaboration for providing the pictures of the KLOE detector and magnet, the Quax Collaboration for the plot of the Q value, and  Enrico Nardi (LNF) and Federico Mescia (Universitat de Barcelona) for their support with the axion preferred-band. We also would like to thank Guido Torrioli (CNR-IFN), Paolo Falferi (FBK), Sandro Gallo and Ruggero Ricci (both LNF) for useful discussions.

\end{document}